# Targeted Nanodiamonds for Identification of Subcellular Protein Assemblies in Mammalian Cells

Short title: Nanodiamond Labels for Biological EM

Michael P. Lake[a], Louis-S. Bouchard[a,b,c,d]*

[a]Department of Chemistry and Biochemistry, University of California, 607 Charles E. Young Drive South, Los Angeles, CA 90095-1569, USA.

[b]California NanoSystems Institute, University of California, 570 Westwood Plaza, Los Angeles, CA 90095-7227, USA.

[c]Department of Bioengineering, University of California, 420 Westwood Plaza, 5121 Engineering V, Los Angeles, CA 90095-1600, USA

[d]The Molecular Biology Institute and Jonsson Comprehensive Cancer Center, UCLA.

*Corresponding author: Louis-S. Bouchard (bouchard@chem.ucla.edu; phone: +1 310 825 1764; fax: +1 310 206 4038)

# Abstract

Transmission electron microscopy (TEM) can be used to successfully determine the structures of proteins. However, such studies are typically done *ex situ* after extraction of the protein from the cellular environment. Here we describe an application for nanodiamonds as targeted intensity contrast labels in biological TEM, using the nuclear pore complex (NPC) as a model macroassembly. We demonstrate that delivery of antibody-conjugated nanodiamonds to live mammalian cells using maltotriose-conjugated polypropylenimine dendrimers results in efficient localization of nanodiamonds to the intended cellular target. We further identify signatures of nanodiamonds under TEM that allow for unambiguous identification of individual nanodiamonds from a resin-embedded, $OsO_4$-stained environment. This is the first demonstration of nanodiamonds as labels for nanoscale TEM-based identification of subcellular protein assemblies. These results, combined with the unique fluorescence properties and biocompatibility of nanodiamonds, represent an important step toward the use of nanodiamonds as markers for correlated optical/electron bioimaging.

# Introduction

Over the past several decades, transmission electron microscopy (TEM) has allowed imaging of biological samples at the single protein level. With recent advances in electron tomography, it is now possible to resolve protein structures in cell sections with sub-nanometer resolution and field of view greater than 1 $\mu m^2$ [1]. Currently, attempts to investigate the structures of proteins *in situ* are hindered by the lack of suitable live cell-compatible labels to identify the protein being observed in TEM. In this study, we show that nanodiamonds (NDs) can be used as targeted intensity contrast labels in biological TEM to image subcellular protein assemblies inside mammalian cells. Nanodiamonds were selected as TEM labels because of their wide range of favorable physical properties for biological applications. Indeed, NDs containing negatively-charged nitrogen vacancy ($NV^-$) defect centers yield strongly fluorescent labels[2] with emission maxima in the near-infrared [3], enabling signal detection *in vivo* from non-superficial tissues [4], and have recently been used to measure temperatures [5] and magnetic fields [6] in living cells with nanometer precision. Nanodiamonds have been used to effectively deliver drugs [7], siRNAs [8], and proteins [9,10], and are currently being evaluated in preclinical trials for delivery of doxorubicin to brain, mammary, and liver tumors [11,12]. Nanodiamonds are also biologically inert and display minimal or no cytotoxicity [2, 13].

In our demonstration of NDs as targeted intensity contrast labels, we have used the nuclear pore complex (NPC) as a model macro assembly. Specifically, we targeted Nup98, a 98-kilodalton nucleoporin and a component of the central nucleoporin ring structure. Nup98 localizes to both the cytoplasmic and nucleoplasmic solvent accessible faces of the NPC and has a critical role in regulating macromolecular import and mRNA export to and from the nucleus [14]. Nup98 is unique among vertebrate nucleoporins and is known to shuttle in and out of the nucleus, trafficking with mRNA to processing bodies (P-bodies) in the cytoplasm and localizing to distinct sites within the nucleus [15,16].

To date, fluorescence microscopy has been the method of choice to determine the subcellular location of functionalized NDs in cells [17,18]. Unfortunately, this method provides a limited view of the cellular environment immediately surrounding the fluorescent probe. However, TEM images of this cellular environment would be desirable to determine what is being measured by the ND. Existing studies of NDs using TEM have been restricted to untargeted or surface modified NDs that remain restricted to endosomes. Previous work predominantly provides isolated snapshots of NDs in cells without quantification [19], with the exception of a recent quantitative study of 150 nm NDs in endosomes [20]. To our knowledge, our TEM analysis is the first quantification of NDs with average diameter under 100 nm selectively targeted to a specific macromolecular structure and outside of endosomes. By utilizing a TEM compatible label [21], we have been able to verify the targeting of NDs to a specific subcellular structure *via* direct visualization.

To unambiguously quantify the localization of antibody-conjugated NDs in mammalian cells at the nanometer scale, we have measured the location of NDs relative to a visible, uniformly sized macro assembly, the NPC. In this study, we have successfully demonstrated the use of maltotriose-conjugated polypropylenimine (PPI) dendrimers to assist in endosomal escape and delivery of NDs to the cytoplasm of HeLa cells, allowing the NDs to traffic for 12 additional hours to the intended target. The goal of this study was to determine if this delivery and targeting method would result in labeling efficiency sufficient to justify use of the NDs to locate a target of interest when visualization of the target may be ambiguous and to determine if cellular factors or the conjugated antibody would dictate the localization pattern instead. Ultimately, this method will assist in low energy, low magnification identification of the NDs, reducing the electron radiation dose on the target of interest during the location stage of TEM imaging.

## Materials and Methods

### Cell Culture

The procedure for transfection, preparation of the ND conjugates, and resin embedding is described in Zurbuchen *et al.* [21] and Mkandawire *et al.* [22].  Briefly, 2 g ($5.7 \times 10^{-5}$ mol) of $4^{th}$ generation PPI dendrimers (SyMO-Chem, Netherlands), 0.92 g (1.82 mmol) and 0.24 ml (1.82 mmol) borane-pyridine complex were mixed in sodium borate buffer at 50° C, stirred continually for 7 days, purified by dialyzing against DI water for 3 days and freeze dried in a lyophilizer.  The nanodiamonds were sonicated in a 3:1 mixture of concentrated $HNO_3:H_2SO_4$ for 24 h to enhance carboxylation of the ND surfaces.

For the conjugated NDs, an antibody raised against Nup98 (C-5), a mouse monoclonal IgG1 from Santa Cruz Biotechnology (sc-74578, Antibody Registry ID: AB_2157953), was conjugated to NDs using 1-ethyl-3-(3-dimethylaminopropyl)carbodiimide (EDC) to covalently attach free amines on the antibody directly to the carboxyl groups on the ND surface.  The antibody:ND ratio during conjugation was 6:1. HeLa CCL2 cells (ATCC CCL-2) [23] were grown on polylysine-treated plastic coverslips for 24 h prior to transfection.  24.3 µg Nup98 conjugated and unconjugated NDs mixed with 2.5 µl Nup98 antibodies with an average diameter of 15-20 nm prepared by ball milling were pre-mixed with 100 µg maltotriose-conjugated PPI dendrimers, incubated at room temperature for 20 min. Transfection reagents were added dropwise to cells in a 6-well dish with 2 ml Dulbecco's modified Eagle's medium (DMEM) without serum at a final concentration of 12.2 µg/ml and 50 µg/ml of ND and glycodendrimer, respectively.  Cells were incubated for 6 h, the media changed to DMEM with 10% fetal bovine serum and then incubated for an additional 12 h post transfection.

## Electron Microscopy Sample Preparation

To obtain material for electron microscopy, materials were directly fixed in a solution containing 2% (v/v) glutaraldehyde and 2% (v/v) paraformaldehyde in 0.1 M phosphate-buffered saline (PBS) buffer (pH 7.4) for 2 h at room temperature and then incubated at 4°C overnight. Subsequently, 0.5% (w/v) tannic acid was added, and the samples were incubated for 1 h at room temperature. The tissues were then washed five times in 0.1 M PBS buffer and postfixed in a solution of 1% (w/v) $OsO_4$ (in PBS; pH 7.2–7.4). The combined treatment with tannic acid, glutaraldehyde, and paraformaldehyde followed by osmification enhanced the staining of membranes. The samples were washed four times in 0.1 M sodium acetate buffer (pH 5.5) and then block stained in 0.5% (w/v) uranyl acetate (in 0.1 M sodium acetate buffer, pH 5.5) for 12 h at 4°C. Subsequently, the samples were dehydrated in graded ethanol (50%, 75%, 95%, 100%, 100%, and 100%) for 10 min per step, rinsed with propylene oxide, and infiltrated in mixtures of Epon 812 and propylene oxide (1:1 and then 2:1 for 2 h each), followed by infiltration in pure Epon 812 overnight. Embedding was performed in pure Epon 812, and curing was done in an oven at 60°C for 48 h. Sections of 80-90 nm and 200 nm thickness were cut with an ultramicrotome (RMCMTX) using a diamond knife and were deposited on copper 200 mesh Quantifoil grids or single-hole grids coated with formvar and carbon, respectively. Sections were then double-stained in aqueous solutions of 8% (w/v) uranyl acetate for 25 min at 60°C and lead citrate for 3 min at room temperature.

## Electron Microscopy

Images were taken on a T20 iCorr cryo-transmission electron microscope from the FEI Tecnai G2 family. Images were taken over wide cell areas at 200 kV using selected area (SA) magnifications ranging from 3,500× to 29,000×, resulting in resolutions from 21 to 3.8 Å/pixel in 2D projections. Image regions were acquired either manually with defocus ranging from −5 to −70 μm or using automated acquisition with Fourier transform-based autofocus over wide areas to obtain unbiased information from each sample. Images were collected on a 2048×2048 electron camera with 16-bit intensity resolution.

## Image Analysis

To assess ND localization statistically, 25 images from the targeted sample and 25 images from the untargeted control sample were selected to be faithful representations of the larger set of images, but without regard to ND quantity or distribution. Each set of 25 images was manually scored to determine edge-to-edge distances from individual ND to individual NPC for all NDs. We processed the images by applying contrast thresholds, morphology-based gradient filtering, and image arithmetic operations as indicated to enhance identification of NDs during the scoring using the Icy bioimage informatics platform [24]. Zoomed images of ROIs were enlarged 10-fold using a bilinear algorithm post-processing for publication.

# Results

## Conjugation of PPI Dendrimers

To assess the effectiveness of antibody-conjugated NDs as labels in a cellular landscape, we have evaluated the spatial pattern of NDs twelve (12) hours after transfection into living cells relative to that of a specific, independently identifiable macromolecular assembly, the NPC. Maltotriose conjugated PPI dendrimers were mixed with NDs prior to transfection to aid in delivery and endosomal escape. Maltotriose conjugation was verified using $^{13}$C and $^{1}$H NMR spectroscopy, and the results were consistent with previously published spectra and descriptions (see Figure. S1, supplementary information) [25,26]. Quantitative comparison of NMR peaks from the $^{13}$C spectrum indicates 83% conjugation of PPI dendrimer free amines (—H$_2$C–NH$_2$) with maltotriose (—H$_2$C–NH$_2$–R, R=maltotriose).

## ND-Nup98 Antibody Conjugates Localize to the NPC

To target NDs to Nup98, anti-Nup98 antibodies were conjugated directly to the surfaces of NDs, and the conjugates were then delivered into the cytoplasm of HeLa cells *via* transfection with 32-branch PPI glycodendrimers to aid in endosomal escape. Twelve (12) hours after transfection, the cells were fixed and processed by standard resin embedding, stained with heavy metals including OsO$_4$ which predominantly binds to membrane regions including the nuclear envelope, and sectioned into ultra-thin slices of 90 nm. Figure 1c shows a higher magnification (29,000×, 3.8 Å/pixel) 2D projection image of a NPC obtained from this preparation revealing the internal structural features including the cytoplasmic ring, the inner ring, the nuclear ring, and the nuclear basket. These features correspond to a recently published structure of the NPC [27] from HeLa cells (Fig 1a) and verify the image scale. A large number of nuclear pores are easily identified in thin sections at magnifications as low as 3400x and appear as clear gaps in the nuclear envelope staining or as visible structures matching the well-defined dimensions of the NPC. Using these features and the ability to detect NDs with high sensitivity, we generated a distribution of the minimum edge-to-edge distances observed from projection images between each ND and the nearest nuclear pore using representative images from the antibody-conjugated sample (targeted) and the unconjugated sample (untargeted) (Fig 2). The experiment was performed from ND conjugation through EM sample preparation 2 times independently and with 3 concentrations of both conjugated and control NDs. Due to lack of saturation at the highest concentration of NDs tested, we selected the highest concentration tested for quantitative analysis.

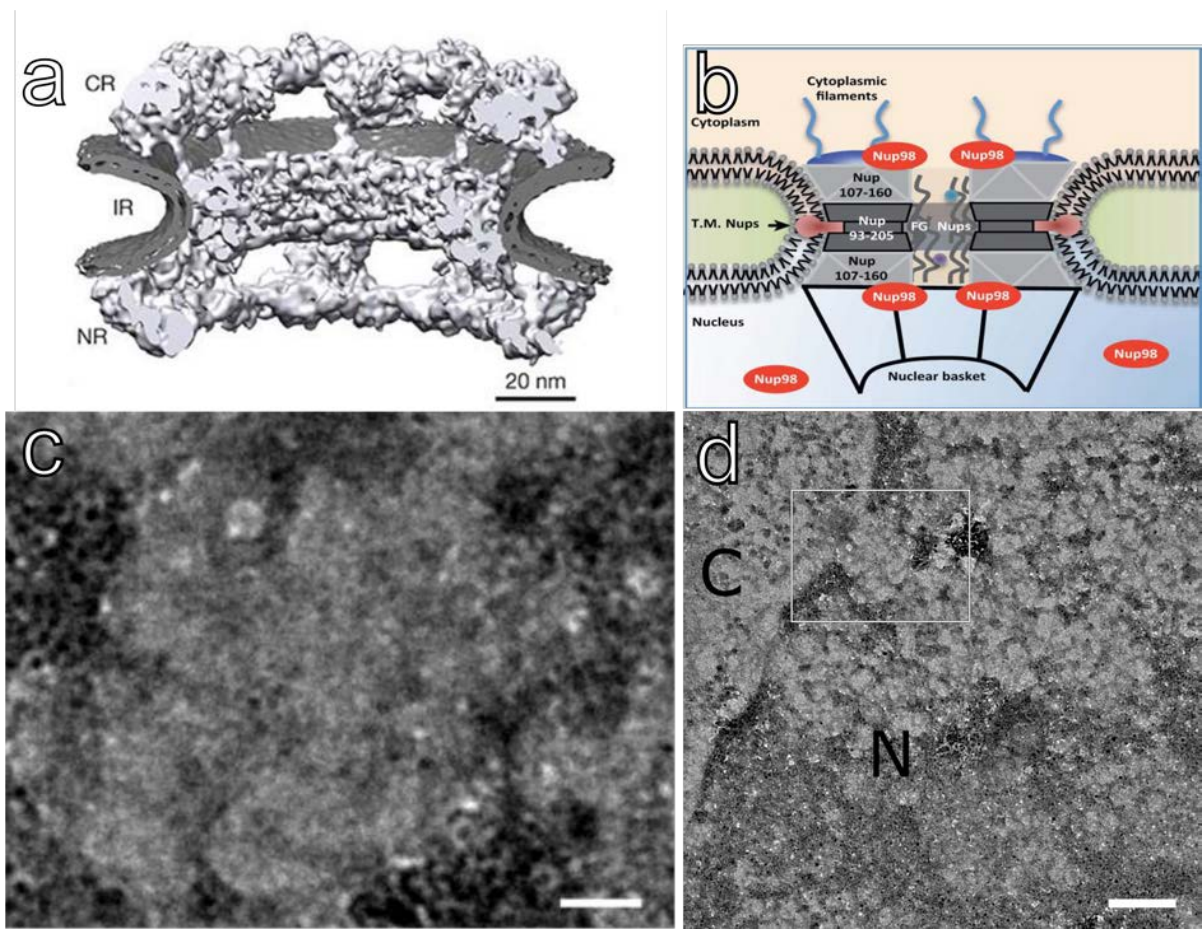

**Fig 1. Nup98 and the NPC as a target for NDs.** (a) *In situ* structure of human nuclear pore from isolated intact HeLa nuclei. Reproduced from Ref. [27] with permission. (b) Location of Nup98

within the NPC showing that Nup98 can localize in the central channel on both sides of the nuclear pore and outside, where it helps anchor the NPC to the nuclear envelope. Reproduced from Ref. [15] with permission. (c) High magnification (3.8 Å/pixel) TEM image of human nuclear pore in 90 nm epoxy resin slice from HeLa cells showing structural details including the cytoplasmic ring, internal ring, nuclear ring and basket. Scale bar, 20 nm, rotated. (d) Wide area view of the NPC in (c), scale bar 100 nm, no rotation. C and N designate cytoplasmic and nuclear regions.

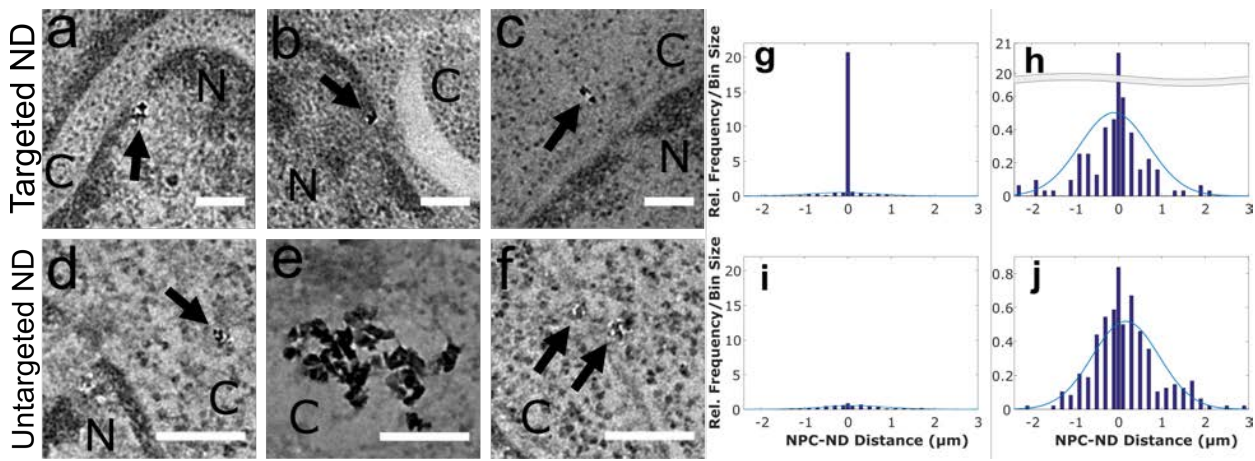

**Fig 2. Successful targeting of NDs to the NPC.** (a–f) Representative images of anti-Nup98–conjugated NDs (a, b) bound to or (c) located near an NPC, and (d–f) unconjugated NDs. C and N designate cytoplasmic and nuclear regions. Scale bar, 125 nm. Black arrows point to NDs. (g,i) Histograms of edge-to-edge distances from each ND or cluster to the nearest nuclear pore for anti-Nup98–conjugated (g) and unconjugated NDs (i). Central bin counts as a percentage of the total NDs counted are 31.0% and 1.26% for (g) and (i), respectively. (h,j) Rescaled data from (g) and (i). Negative and positive values refer to distances into the nucleus and cytoplasm, respectively. We analyzed twenty-five (25) representative images from both the conjugated and unconjugated TEM samples and identified 158 and 239 NDs or ND clusters, respectively, including approximately 1,000 intracellular NDs for both sets. Minimum distance (bin 0) was defined as ± 7.5 nm to account for antibody length. The center 3 bin edges are: -200, -7.5, 7.5, 200, and all others increment by 200 nm. The blue curve is a normal distribution based on the mean and standard deviation of the targeted NDs not bound to the NPC and the untargeted NDs, respectively. The distances measured are included in the Supplementary Information as Table S2.

We selected twenty-five (25) representative images from both conjugated and unconjugated (control) ND samples from over 100 images acquired for each sample and scored them manually to determine the distribution of edge-to-edge distances. An example of the scoring is provided in the supplementary information (Figure S3). Image selection was blind to the position of any image inside of the larger image set. Automatic focus and image acquisition over pristine, random areas was used to collect images without bias, with prescreening only to identify areas that contained cells. To construct a binary statistic, we categorized NDs within 7.5 nm of either side of the pore as bound to account for the size of an antibody and the location of Nup98 at the edge of the NPC, represented by the central bins in Fig 2, panels (g) and (h). Using this criterion to define the bound fraction, 31.0 percent of all intracellular anti-Nup98–conjugated NDs or clusters were bound to a NPC, compared with 1.26 percent for unconjugated ND or clusters. ND counts per cluster had a mean $\mu = 9.4$ and a standard deviation $\sigma = 11$ for the targeted NDs and mean $\mu = 7.5$ and standard deviation $\sigma = 6.3$ for the untargeted NDs. These values were consistent across several counts. We also analyzed the images with a different binary statistic by defining each NPC as labeled or unlabeled using the same distance criteria (within 7.5 nm), resulting in 32/89 or 36.0% NPCs labeled by NDs in the targeted set and 3/93 or 3.2% NPCs labeled in the untargeted set. Treating the probability of an NPC being bound to a ND as a Poisson statistic, we then performed a test of the null hypothesis that these counts were generated by the same underlying probability of success for unequal sample size as described by Shiue and Bain [28].

Briefly, for two Poisson distributions $X \sim POI(\lambda_1=S_1*\gamma_1)$ and $Y \sim POI(\lambda_2=S_2*\gamma_2)$, where S is the total number of observations and $\gamma$ is the probability of success, the conditional distribution of Y, given the counts observed are Y = y and (X + Y) = x + y = m, is a binomial distribution with parameters m and p or $(Y | x + y = m) \sim Bin (m,p)$ where $p = S_2/(S_1+S_2)$.  This conditional distribution provides a universally most powerful unbiased (UMPU) test of the hypothesis $H_0$: $\gamma_1=\gamma_2$ against the alternative $H_a$: $\gamma_2 > \gamma_1$ at the level of $1 - Bin(y-1; m,p)$.  To use this test, we observe y = 32 (targeted), x = 3 (untargeted), x + y = 35 and determined that 1-Bin(31; 35, 89/182) = $1.1 \times 10^{-7}$, or 1 – the cumulative probability that y ≤ 32, allowing us to reject the null hypothesis for any alpha $\geq 1.1 \times 10^{-7}$.  Similar analysis taking the counts of central bin of the above histograms as successes in a Poisson distribution gives y = 49, x = 3, and x + y = 52 yielding 1-Bin(48; 53, 158/397) = $1.5 \times 10^{-14}$. This information is summarized in the supplementary information Table S1.

Comparing the untargeted data to a normal distribution using a chi-square goodness-of-fit test we determined a p-value of 0.027, which indicates a reasonable deviation from a normal distribution, especially considering that this includes both cytoplasmic and nuclear distances in the same distribution.  Assessing the targeted data set by excluding the central bin and assessing all remaining values using the same test, we obtain a p-value of .17, indicating that the distance from nuclear pores of the unbound fraction of targeted diamonds are close to normally distributed. Due to manual identification of nuclear pores it is possible that the closest NPC is not detected, however identification of the nuclear envelope is unambiguous is all images and given the high density of nuclear pores found on the nuclear envelope compared to the distances being measured, an error in identification would only slightly alter the distributions presented.

## Anti-Nup98–Conjugated NDs as Markers for the Nuclear Pore

Anti-Nup98–conjugated NDs bind to the NPC in a manner that reflects the expected localization pattern of Nup98 (Fig 3, a–e). The NDs tend to appear in clusters centered in the pore or off-center, consistent with having been dislodged and left at the edge of the NPC on either side of the membrane, possibly due to interference by endogenous molecular traffic. Considering that the NDs in this study have a size (~20 nm) smaller than the NPC central channel (~60 nm) [27], aggregates of NDs within the pore are not unexpected. Nup98 is known to localize to other cellular structures, and thus the calculated percentage of conjugated NDs localized at the NPC (31.0%) likely represents a lower bound for the targeting efficiency of anti-Nup98–conjugated NDs. Consistent with this, we observed strong localization of NDs to additional distinct structures. Identifying and properly quantifying binding to these other structures is hindered by lack of positive identification, however, and will be the subject of future investigations. Fidelity of the ND conjugates for Nup98 overall appears significantly higher than the percentage reported bound to NPCs, but can only be qualitatively assessed by observation at this time. It is known that Nup98 localizes to multiple subcellular assemblies such as P-bodies and intranuclear bodies however, these structures cannot be unambiguously identified in our TEM images. Most ND localization studies with larger NDs do not observe ND entry in to the nucleus. This claim is not consistent with our observations with ~20 nm NDs for either targeted or untargeted NDs in this study. However, our study was not intended to analyze this characteristic and future systematic studies of this observation coupled with additional measurements should be more conclusive. It has been observed that 4 nm gold particles modified with both and poly(ethylene glycol) [29] and polyethylenimine [30] and 70 nm $SiO_2$ nanoparticles [31] form aggregates in the nucleus, emphasizing the importance of further research on the ability of NDs to enter the nucleus.

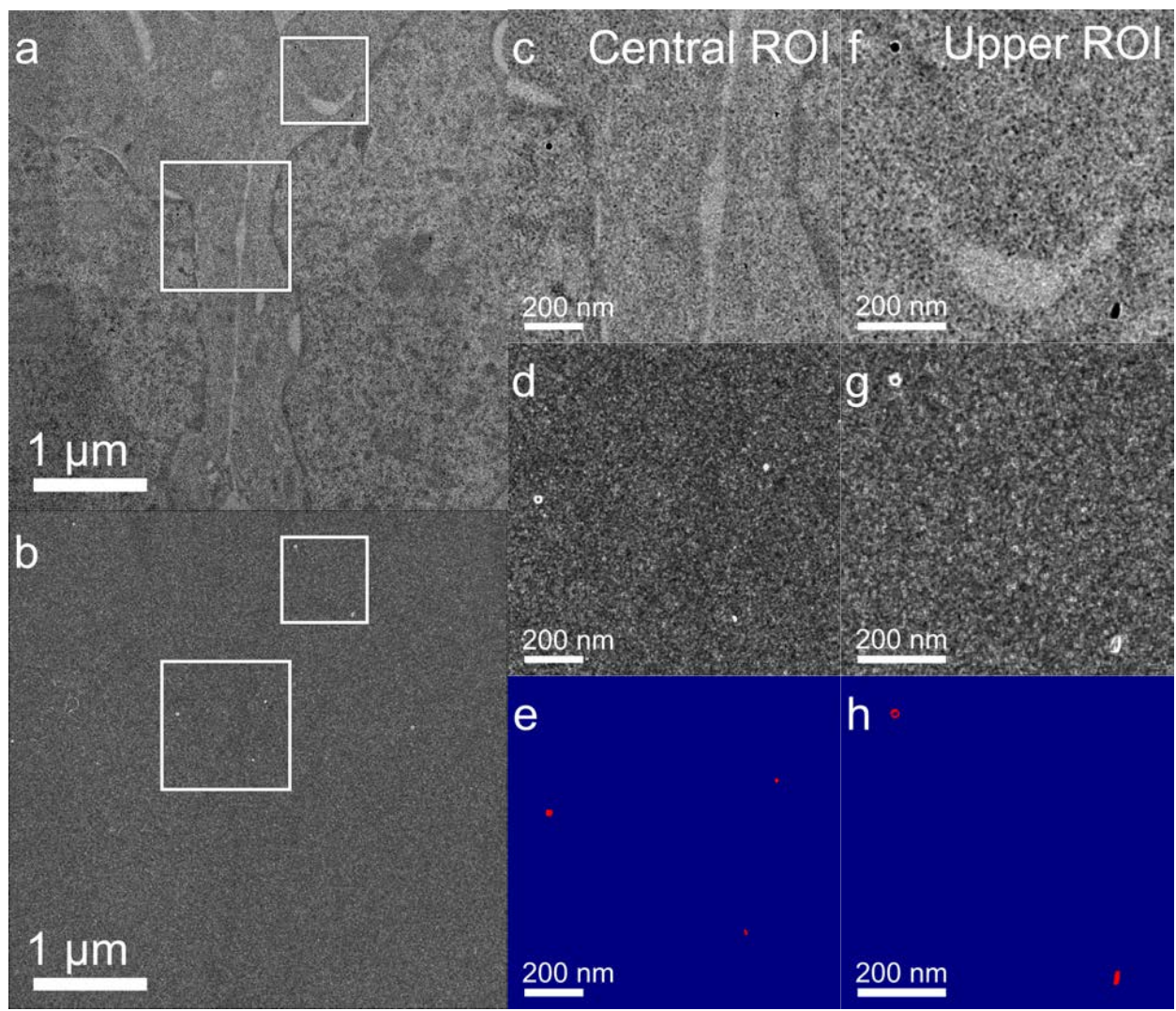

**Fig 3. Nanodiamonds can be detected individually from low magnification (5,000×) images.** (a) Original image. (b) Zoom of ROI from (a). (c) Gradient morphometry enhancement of (b). (d) Contrast inverted image of (b). (e) Inverted and gradient images summed with threshold applied to isolate diamond. Red and blue indicate saturated maximum and minimum intensities, respectively, resulting in a binary colored image.

## Isolating NDs from Cellular Background

To further assess the utility of NDs as TEM landmarks, we evaluated the intensity profile of NDs in transmission electron images. We found that the NDs are clearly distinguished from both the surrounding intracellular milieu and heavy metal stained components because of their low transmitted intensity and bright intensity fringes. Figure 3 provides an example of these features in raw and processed images captured under low magnification, demonstrating that they can be used to detect NDs with zero background. The NDs were successfully isolated from the cellular background in heavy metal-stained samples using a simple two-step image processing method. First, a gradient morphometry filter [24] was used to generate a separate image. Then, by inverting the intensity of the original image to turn dark contrast into bright signal and superimposing the two images, a clearly distinguishable object for each ND results. NDs are detected with high sensitivity because they appear as simultaneously bright and dark objects on the scale of tens of nanometers, producing an exceptionally strong spatial gradient, in addition to their intensity contrast. It can be seen from Fig 4 that NDs can be isolated with zero background signal over areas spanning significant fractions of cells that include cytoplasmic, nuclear and extracellular regions. Because the appearance of NDs in TEM images depends on the crystal orientation and the defocus, most (but not all) NDs observed can be isolated with zero background from automatically acquired images with human intervention only to select coverage areas. Based on the results from this acquisition method, alteration of the acquisition process to include a limited series of tilt angles or defocus values and diffraction mode imaging at each location should enable all individual NDs to be isolated with zero background in wide area images. Our previous work has demonstrated that diffraction-mode imaging can identify NDs over a magnification range spanning seven orders of magnitude, and that high resolution TEM can conclusively verify the identity of NDs by observing the atomic lattice and by generating diffraction patterns [21].

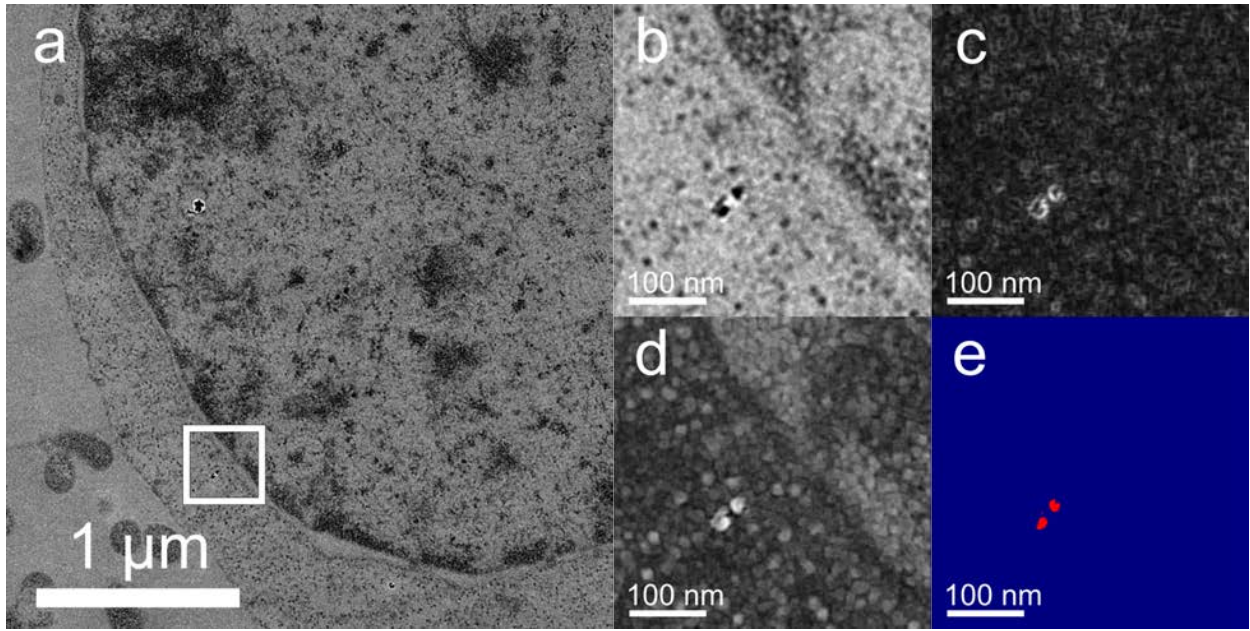

**Fig 4. Nanodiamond isolation is robust and reproducible.** (a) Wide area view of two cells transfected with anti-Nup98–conjugated NDs. (b) Gradient enhanced image of (a). (c) Central ROI from (a). (d) Central ROI from (b). (e) Original image inverted and added to gradient image, then contrasted to show isolation of ND with zero background. (f) Upper ROI from (a). (g) Upper ROI from (b). (h) Original image inverted and added to gradient image with threshold applied to show isolation of ND with zero background.

In epoxy resin-embedded samples stained with $OsO_4$, the predominant source of membrane contrast is the osmium, which concentrates in lipid membranes like the nuclear envelope. Because of this, objects near the nuclear membrane are especially challenging to isolate. Figure 5 shows complete isolation of the NDs that are bound to the nuclear pore, with substantial sections of the nuclear membrane in the field of view. Because this data was collected with the most basic single image acquisition method and a simple image processing algorithm, there is substantial opportunity to enhance the specific detection of NDs over wide areas in a biological environment to achieve perfect fidelity for ND recognition.

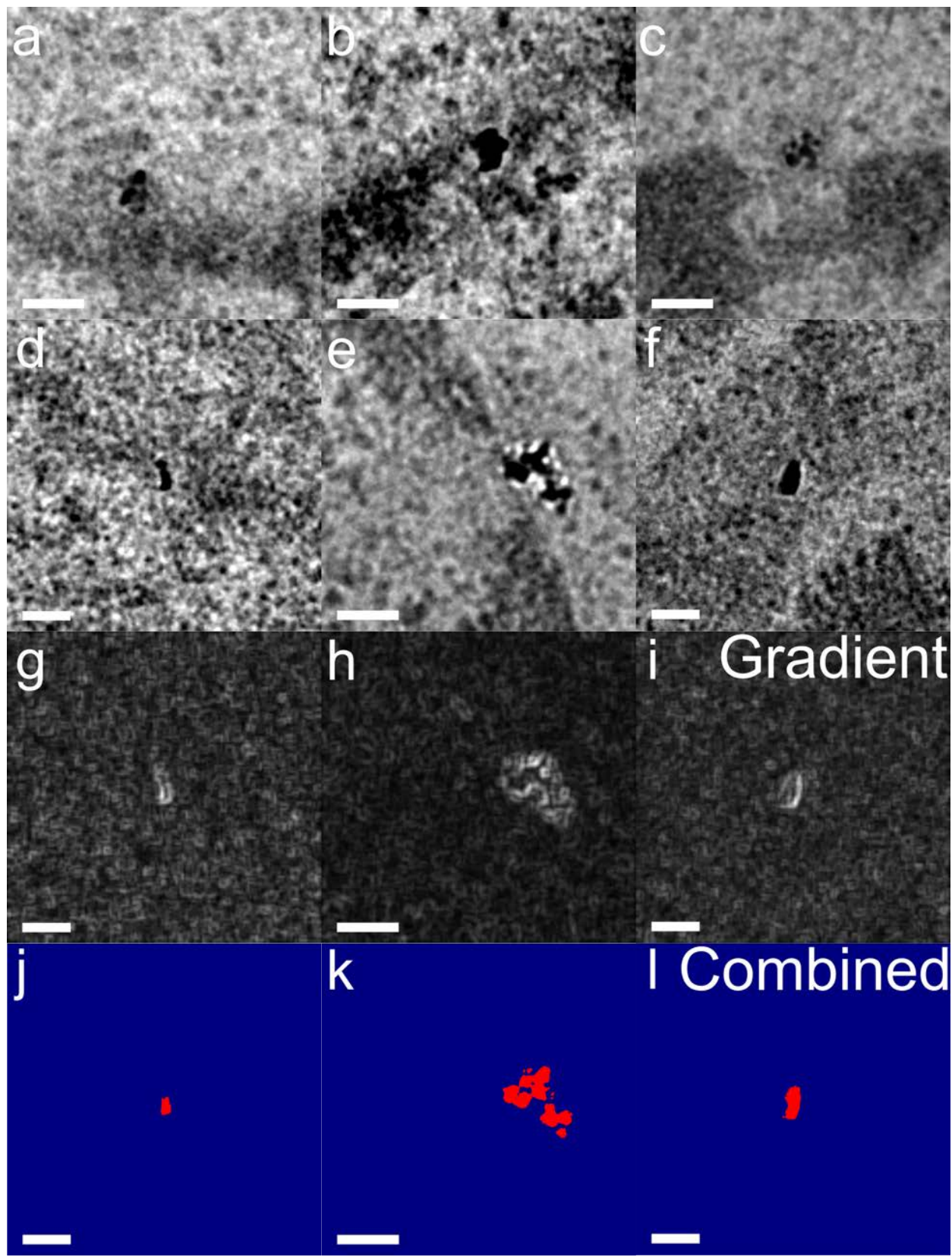

**Fig 5. Isolation of NDs bound to NPCs.** (a–f) Unprocessed TEM images of anti-Nup98–conjugated NDs bound to NPCs in 90 nm epoxy resin sections from HeLa cells fixed 12 h after transfection with the ND conjugates. NPCs are highlighted in green. (g–i) Gradient morphometry filter applied to (d–f). (j–l) Images processed with a color threshold applied to isolate NDs. All scale bars are 50 nm.

Due to their geometry and crystalline volume, NDs produce bright thickness and phase fringes that make them excellent candidates as fiducial markers for computer recognition during tomographic reconstruction and computer-assisted localization. Figure 6 shows that the gradient image alone is sufficient to isolate most individual NDs. From the ND gradient image (center column), the image-based signal-to-noise ratio (SNR) of the ND center to fringe transition, given as the mean intensity of the brightest circle of pixels of the ring divided by the standard deviation of the local background are 14.0 and 11.3 (15,555/1,044 and 11,791/1,041) with the brightest pixel in each having an SNR of 18.3 and 14.4, for panels (a) and (b), respectively. Because the fringes derive mainly from the shape of the nanocrystal [32], the ring-shaped fringe indicates that the particles are quasi-spherical. The development of consistently shaped NDs from processes other than ball milling should result in reproducible signals such as the two shown in Fig. 6, significantly enhancing the detectability of NDs in cells using TEM.

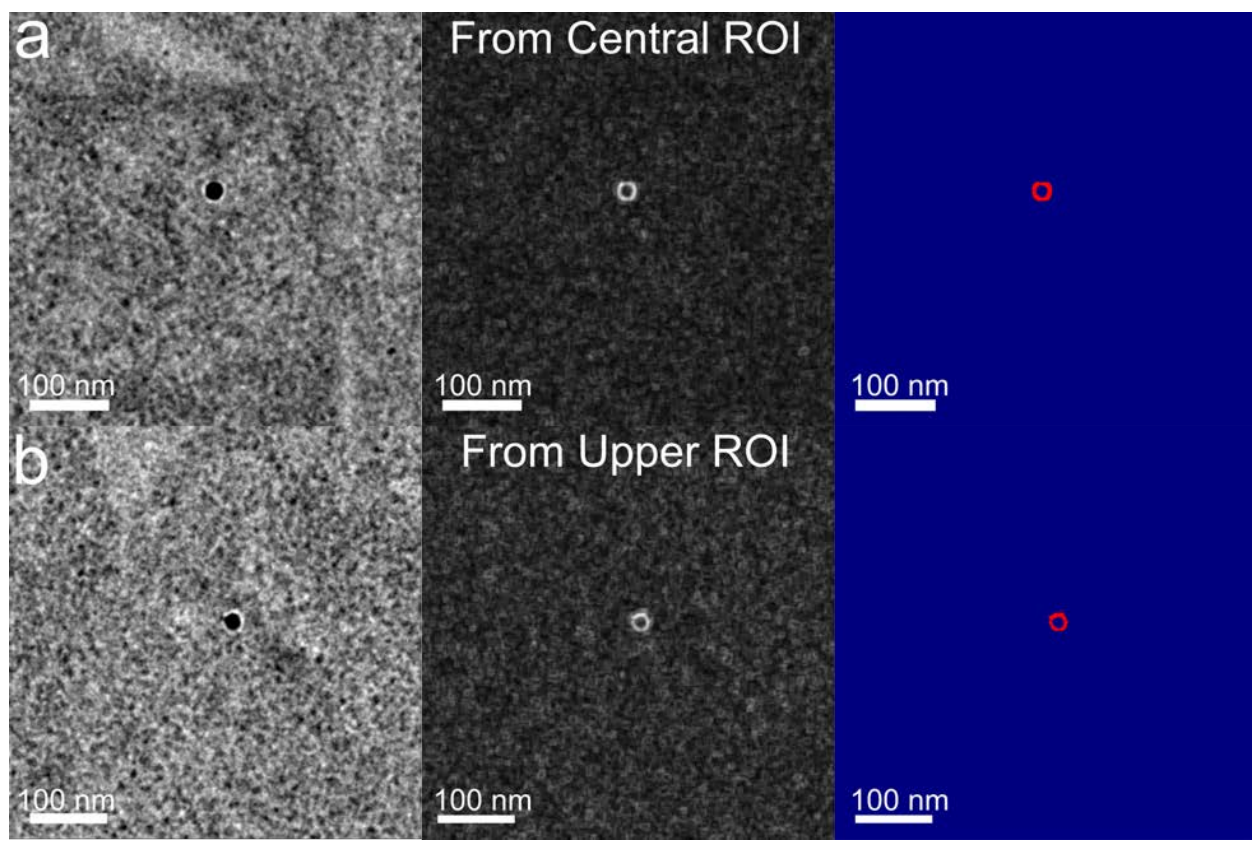

**Fig 6. Gradient enhancement alone allows detection of quasi-spherical NDs.** Cropped sections from (a) central ROI and (b) upper ROI of Fig. 4. From left to right are the original images (left), gradient morphometry enhanced images (center), and contrast thresholds applied to the gradient images (right).

## Discussion

Fluorescent labeling of proteins has been highly successful for sub-micron visualization of protein localization. Nonetheless, the resolution of optical measurements is subject to Abbe's diffraction limit [33] and super-resolution techniques do not allow for visualization of the cellular environment surrounding fluorescent labels at the nanometer scale. Fluorescence microscopy studies indicate that the cellular uptake mechanism for unconjugated NDs is endocytosis and that NDs are subject to several cellular trafficking mechanisms [19].

Although fluorescence microscopy can determine the localization of particles to the perinuclear region, more advanced optical techniques and carefully controlled experiments are required in order to differentiate targeting to more specific cellular regions, such as localization to the nuclear *v.* cytoplasmic sides of the nuclear envelope [14]. Moreover, the diffraction limit of light (~200 nm) imposes a significant obstacle to the mapping of 20 nm nanoparticles particles relative to a 120 nm macro assembly such as the NPC. Light microscopy precludes discrimination among particles that are bound to the intended target, that are near the target, that have displaced the target protein from its assembly, or particles that are bound in aggregates or nanoscale structures preventing them from directly contacting the intended target.

While sub-diffraction methods such as STED microscopy can visualize fluorescent molecules such as $NV^-$ NDs, these methods are limited by substantial sample restrictions, the requirement for high laser powers that can bleach fluorophores and damage intact cell samples, inability to distinguish large NDs from clusters because they do not detect the diamond material and the requirement to co-stain any target, severely limiting the interpretation of the context surrounding the target and label. In practice, most sub-diffraction imaging systems are restricted to 40-80 nm resolution in cells [34], especially because STED resolution depends strongly on the power of the depletion beam. Furthermore, even ~20 nm resolution is insufficient for structural identification of individual macromolecular assemblies such as the nuclear pore or to separate individual NDs when clustered, precluding accurate quantification [35]. In fact, this methodology was specifically inspired as a way to complement STED and other fluorescent imaging techniques, by providing much of the information inaccessible through these methods.

Nanoscale localization information about the Nup98 nucleoporin of the NPC by immuno-gold staining of cells post fixation has been studied [36]. However, the ability of antibodies to successfully localize NDs to subcellular regions with nanoscale precision and outside of endosomes in live cells has not been demonstrated prior to this study. Given the resolution limit of light microscopy and the emerging use of NDs as nanoscale optical sensors in live cell thermometry and magnetometry, as well as in drug and gene delivery, knowledge of the precise location of NDs relative to a target of interest is increasingly important.

In previous work, we demonstrated that NDs possess the physical and biological properties necessary to act as landmarks in a cellular environment for TEM using resin-embedded ultra-thin sections from HeLa cells [21]. Whether the localization of antibody-conjugated NDs to specific protein assemblies is sufficiently robust to enable live cell labeling of protein assemblies, accounting for transfection and trafficking throughout cells, has not been previously determined and is the subject of this work. In general, quantification of the subcellular location of nanoparticles is difficult due to the large number of sections and images required to unambiguously identify nanoparticles, usually at higher magnifications than the present study, and the difficultly of normalizing to the variable geometry or volumes of the sub cellular regions being compared. [37, 38] By utilizing a recognizable, homogeneously sized macromolecular structure, the NPC, and the binary criteria of being inside or outside of the nucleus, our analysis avoids these complications.

We show here that NDs provide a robust, identifiable landmark that can be used to direct expensive and time-consuming tomography studies to specific regions of interest (ROIs), providing a high probability of detecting the target of interest, even if the identity of the protein assemblies and their constituents cannot be verified until after image processing and reconstruction. Using this method, low magnification images with wide fields of view can provide quantification of NDs or identification of ROI's around NDs with automated image collection.

Low SNR is an inherent limitation in TEM of cells; increasing beam exposure enhances SNR at the cost of increasing damage to the sample.

Targeting of biologically inert contrast agents such as NDs with high specificity to subcellular structures can extend the applications for TEM imaging of biological structures by providing a multi-functional label capable of correlating fluorescently detected sensor signals and images with structural information obtainable exclusively by TEM.  Nanodiamonds offer several distinct advantages over other nanoparticles used for probing biological systems using TEM.  Quantum dots (QDs) are amenable to surface functionalization but lack the versatility of binding, especially directly to proteins, offered by a heavily carboxylated surface.  They are also generally cytotoxic [39, 40] and are prone to intermittent fluorescence [41].  Moreover, QD fluorescence is quenched by osmium tetroxide fixation [42] limiting the practicality of QDs for multimodal imaging.  Gold nanoparticles (AuNPs) lack the biosensing capabilities of NDs [5], do not readily undergo endosomal escape, accumulating in endosomes even after accounting for diverse strategies to aid in cytoplasmic delivery [43,44], and are not easily distinguished by appearance in electron micrographs from $OsO_4$-stained cytosolic lipid droplets [45].

Unlike NDs, AuNPs are not bioinert and produce cytotoxic effects at concentrations as low as 20 nM [46].  AuNPs may also induce protein aggregation at physiological pH due to the high affinity of gold for free thiols on proteins, which are abundantly present in the reducing environment of the cytoplasm [47].  While capping and surface modifications can reduce these effects, their removal *in vivo* and resulting exposure of the bare gold surface makes modified AuNPs unattractive for long-term biological labeling or clinical applications [48]. Finally, metal oxide nanoparticles are useful as contrast agents in TEM but exhibit high cytotoxicity [49,50] and lack the versatility of NDs, especially the sensing capabilities and the availability of surfaces for direct covalent conjugation of proteins.

Based on our results, targeted NDs can identify specific sub-volumes or cellular ROIs with sufficient efficiency to target high-resolution imaging or electron tomography. To improve detection, acquisition of a minimal defocus series including positive defocus at each location will enable detection of variations in ND fringes with the defocus, further capitalizing on the material differences between crystalline and amorphous carbon structures, which ideally show no contrast at positive defocus. Furthermore, this signature can be analyzed using image-based, high-content recognition algorithms similar to those currently used in fluorescence microscopy to identify subcellular geometric patterns in drug screening assays [51]. Automating the recognition of fiducial or correlation markers over whole cell areas provides a major advantage for *in situ* imaging approaching the structural level, fiducial marker identification for reconstruction of electron tomograms and single particle reconstructions of large numbers of repetitive images to refine structures. This method has direct applications in traditional TEM as well as in focused ion beam thinned cryo sections, where the imaging necessary to locate structures of interest can significantly degrade the sample. We are currently optimizing methods to extend this technique to correlated light (live cell) and electron microscopy without the need for specialized microscopes or stages.

Beyond research, this method has applications for improving the analysis of both ND and monoclonal antibody-based therapeutics. TEM-based ND imaging enables the localization of therapeutic antibodies bound to NDs in tissues of humans or animals to be determined unambiguously at the nanometer scale. To date, an increasing array of monoclonal antibodies have been approved by the Food and Drug Administration (FDA) to treat cancer, autoimmune diseases, allergic asthma, viruses, blood disorders, organ transplant rejection, osteoporosis, and several other disorders [52]. PPI dendrimers also have promise in therapeutic applications.

*In vivo* studies with rats have demonstrated that 25% conjugation with maltotriose reduces cytotoxicity to a minimal level and 100% conjugation eliminates detectable toxicity at all doses tested [25]. When tested on panels of human cells for pharmaceutical applications, densely maltotriose-conjugated dendrimers were the least cytotoxic form of PPI glycodendrimers, compared to unconjugated, partially conjugated and maltose-conjugated PPI dendrimers [53]. Given their low cytotoxicity profile, maltotriose-conjugated dendrimers hold promise as a component of therapeutics for use in humans [54]. The third part of our targeting strategy, the NDs, are also generating increasing interest as therapeutic agents. *In vivo* studies in mice have shown that intraperitoneal injection of NDs did not cause detectable cytotoxicity in the central or peripheral nervous system or impair neural function based on gross animal behavior and hippocampal novel object recognition (NORT) tests which can report damage to the hippocampus [55].

A recent study concluded that 20 and 100 nm carboxylated NDs showed no cytotoxicity or genotoxicity measured in real time in six human cell lines derived from liver, kidney, intestine and lung tissue, with ND doses up to 250 μg/ml and recommended that carboxylated NDs be considered for use as a negative control in nanoparticle toxicity studies [13]. The added advantage of detectability by TEM at the nanoscale compared to other candidate medicinal nanoparticles or antibodies alone could improve and expedite clinical trials for NDs as vehicles for antibodies, proteins, nucleic acids or drugs, potentially lowering costs and reducing ambiguity about the subcellular fate of these particles in preclinical studies.

Due to the long-term chemical stability of NDs, this imaging method could also help assess tissue penetration and persistence of therapeutic NDs in biopsies or post mortem tissue immediately following or months after ND administration. Nanoscale imaging of NDs identified as bound to or near a target of interest or accumulating inside cellular structures where they are isolated from the target of interest can correlate therapeutic efficacy measurements with ND location by direct visualization.

Determination of the mechanisms of action and drug resistance for monoclonal antibody therapies are increasingly important for improving clinical outcomes and studies on trastuzumab emtansine show that acquired resistance may be due to reduced binding of the antibody to cancer cells, inefficient internalization or intracellular trafficking, heightened recycling of the antibody-target complex or impaired lysosomal degradation [56]. The ability to successfully target and identify the subcellular location of ND-antibody-target complexes by TEM could contribute significantly to further studies of these mechanisms.

## Conclusions

We have demonstrated a method to prepare, deliver and image ND-antibody conjugates for labeling of cellular protein assemblies. We chose the Nup98 nucleoporin, a component of the NPC, as the target for our study because of the recognizable structure of the NPC, its well defined location and the significance of its physiological functions for sensing and drug delivery applications. Using TEM, we assessed the percentage of NDs bound to the Nup98 nucleoporin at the NPC on a single diamond–single complex basis. In addition to localization at the NPC, we observed clustering of diamonds at other locations, distinct from the NPC, where Nup98 is known to localize. Our results demonstrate that NDs can be identified individually over the area of entire cells at low magnification with near zero background signal from first pass automated image acquisition.

Using the unique signature of NDs observed under TEM in biological samples, we have measured the distance of individual NDs to individual NPCs from images taken over wide areas and with automated acquisition, demonstrating the minimal image requirements necessary to identify NDs. Given the established efficacy of $NV^-$-containing NDs for fluorescence microscopy applications, our results highlight the value of NDs for correlated imaging. Widespread use of correlated fluorescence and electron imaging has been hindered by practical limitations surrounding the markers and sample preparations. Here we have described a TEM-based method that enables high precision localization of particles and the ability to recognize ROIs with a novel TEM marker. This method is compatible with live cell optical microscopy using an established non-bleaching fluorophore ($NV^-$ ND), allowing for practical correlated imaging that can be performed on separate light and electron microscopes, thereby facilitating extension of correlated imaging beyond specialized instruments to equipment readily available at most research institutions. Our technique has the ability to complement and outperform the common method of fluorescence-based localization. We have recently extended this technique into plunge-frozen, whole, unstained cells to provide a versatile and unambiguous standard that can provide localization of particles at the level of proteins and protein assemblies without requiring costly embedding and sectioning.

We further investigated the appearance of NDs observed by TEM, and noted that the maximum and minimum transmitted intensities from NDs are each more extreme than the signal variation from the resin-embedded cellular background or heavy metal staining. We also observed geometrically consistent, ring-shaped intensity gradients likely due to a combination of thickness-and phase-based intensity fringes, capable of providing well defined geometric features for computer-based recognition. These features will likely prove valuable for automated acquisition and tomographic reconstruction.

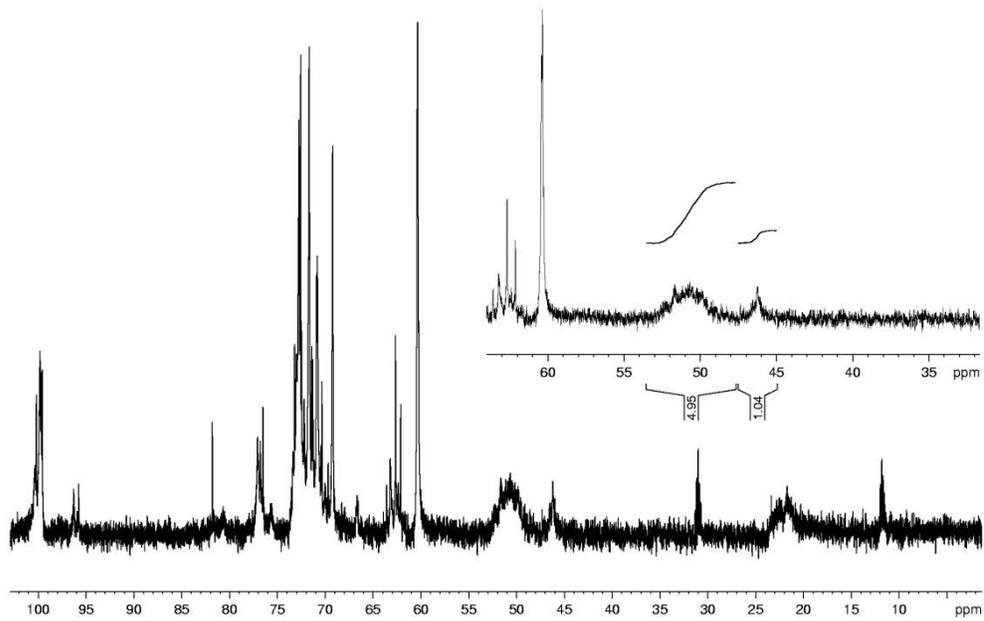
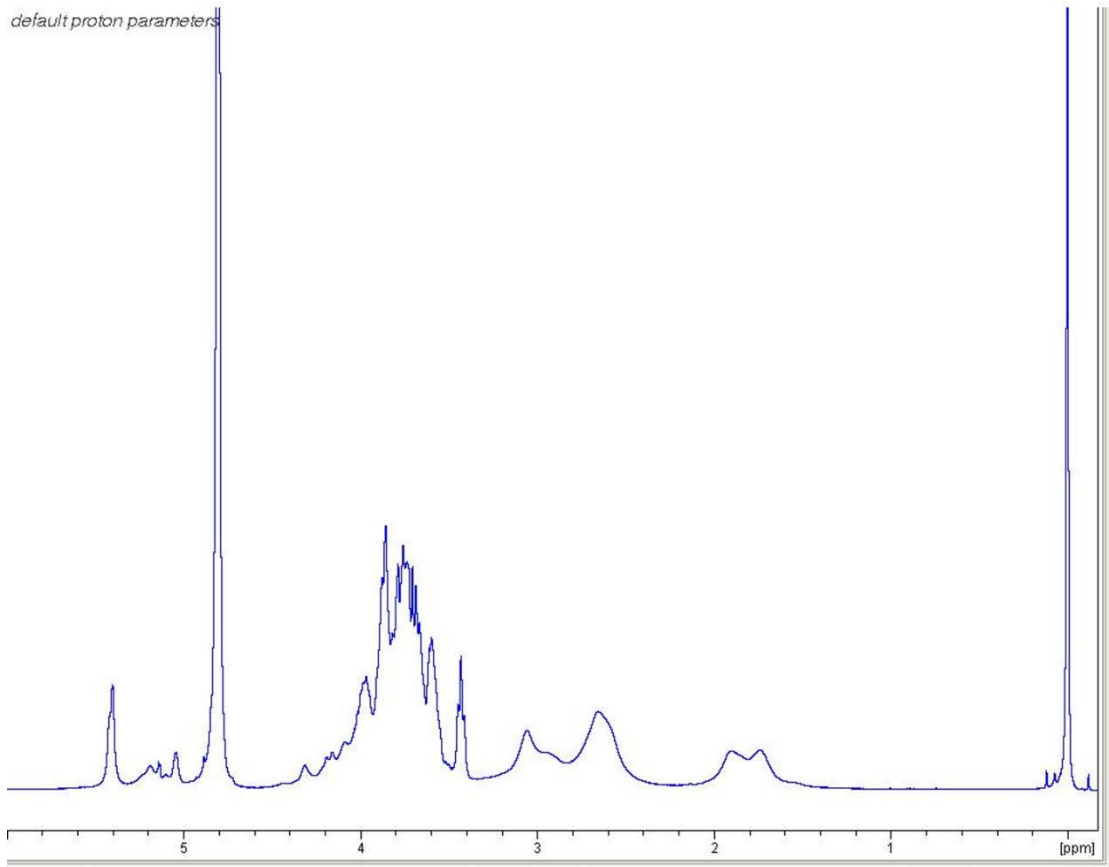

**Figure S1. NMR Measurement of PPI Dendrimer Conjugation.** $^{13}$C (top) and $^{1}$H (bottom) spectrum of maltotriose-conjugated PPI dendrimers recorded on Bruker AVX-500 and DRX-500 NMR systems, respectively. Quantitative comparison of NMR peaks from the $^{13}$C spectrum indicates 83% conjugation of PPI dendrimer free amines (—H$_2$C–NH$_2$) with maltotriose (—H$_2$C–NH$_2$–R, R=maltotriose).

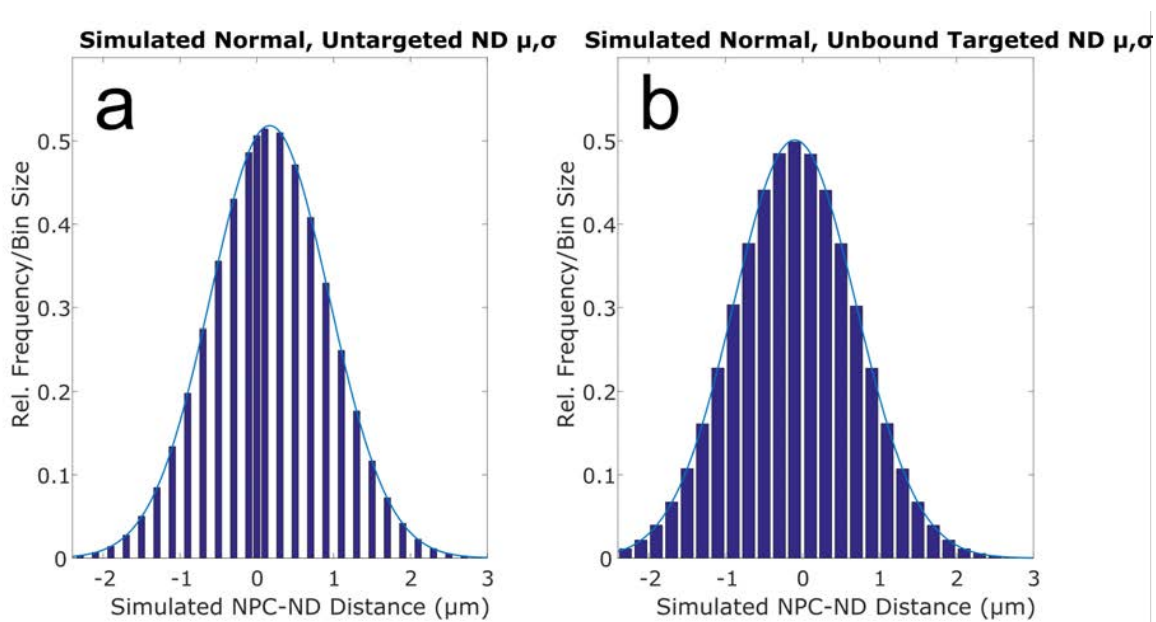

**Figure S2. Impact of Binning Choice on Distribution Presentation.** (a) Histogram of simulated counts to confirm that the relative frequency/bin size reproduces the normal distribution for the untargeted ND data. Data was simulated by generating a 10,000,000 element random data set from the normal distribution fit to the untargeted ND data set (μ = 0.168 μm, σ = .770), then binned identically. Bin edges are -2.4, -2.2, -2, -1.8, -1.6, -1.4, -1.2, -1, -0.8, -0.6, -0.4, -0.2, -0.0075, 0.0075, 0.2, 0.4, 0.6, 0.8, 1, 1.2, 1.4, 1.6, 1.8, 2, 2.2, 2.4, 2.6, 2.8, 3. Counts were then divided by 10,000,000 and this relative frequency was divided by the bin width. (b) Histogram of simulated counts to confirm that the relative frequency/bin size reproduces the normal distribution for the unbound targeted ND data. Data was simulated by generating a 10,000,000 element random data

set from the normal distribution fit to the unbound targeted ND data set ($\mu$ = -.101 $\mu$m, $\sigma$ = .796). This data set was the targeted ND distances with the bound fraction removed consisting of 108. All 10,000,000 elements were binned identically to the entire targeted ND data set. Bin edges are -2.4, -2.2, -2, -1.8, -1.6, -1.4, -1.2, -1, -0.8, -0.6, -0.4, -0.2, -0.0075, 0.0075, 0.2, 0.4, 0.6, 0.8, 1, 1.2, 1.4, 1.6, 1.8, 2, 2.2, 2.4, 2.6, 2.8, 3. Counts were then divided by 10,000,000 and the central bin (bound fraction) was discarded.

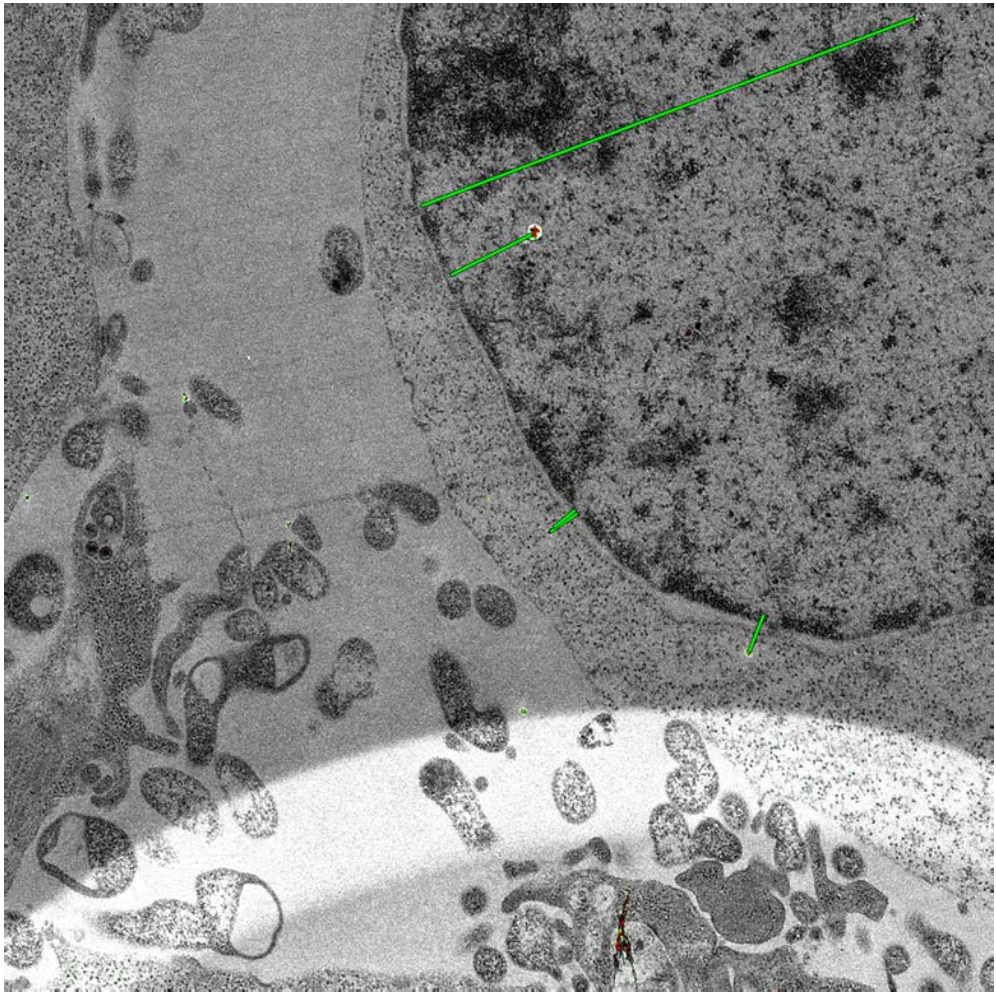

**Figure S3. Scored Targeted ND Image.** Example of image with scoring from targeted ND image set. Only intracellular NDs were included.

**Table S1. Poisson Statistics for Labeling of NPCs**

**Table S2. NPC-ND Distances Measured from TEM Images.**